\documentclass{article}
\usepackage{slashed,amsmath,amssymb,enumitem}
\usepackage[papersize={8.5in,11in}]{geometry}
\begin{document}
\title{Stochastic Quantum Gravity, Gravitational Collapse and Grey Holes}
\author{J. W. Moffat\\~\\
Perimeter Institute for Theoretical Physics, Waterloo, Ontario N2L 2Y5, Canada\\
and\\
Department of Physics, University of Waterloo, Waterloo, Ontario N2L 3G1, Canada}
\maketitle
\begin{abstract}%
Quantum gravity is treated as a stochastic quantization of the spacetime metric of general relativity.  It is found that
owing to the stochastic fluctuating behavior of the geometry, the singularity in gravitational collapse of a star
has a zero probability density of occurring. Moreover, as a star collapses the probability density of a distant observer seeing an infinite red shift at the Schwarzschild
radius of the star is zero. Therefore, there is a vanishing probability density of a black hole event horizon forming during gravitational collapse. A grey apparent event horizon forms 
during the collapse that allows information to escape.
\end{abstract}



\section{Introduction}
In 1996-97 I proposed that quantum gravity can be described by a stochastic quantization of the spacetime metric of generally relativity (GR)~\cite{Moffat}. A general covariant method for stochastic quantization of the gravitational spacetime metric was formulated.  It was demonstrated that the probability density for a singularity to occur in the collapse of a star was zero and that a classical black hole event horizon is not produced in the collapse.  Moreover, the probability density for the light emitted from the surface of the collapsing star to have infinite redshift is zero. A grey apparent horizon is formed during the collapse and the fluctuations of the surface allow for the escape of light avoiding the information loss paradox~\cite{Hawking,Hawking2,Hawking3}.

Stephen Hawking has proposed a resolution of the information loss paradox by claiming that gravitational collapse does not produce a classical black hole event horizon behind which information is lost~\cite{Hawking4}. Instead, it produces an apparent horizon that like weather forcasting on Earth retains quantum unitarity and an effective loss of information. Hawking avoids the need for a ''firewall" that has been suggested to occur at a black hole event horizon~\cite{Polchinski}. By incorporating quantum mechanics into GR, Hawking has demonstrated that a black hole radiates with a black body thermal spectrum~\cite{Hawking,Hawking2,Hawking3}. Near the event horizon of a black hole a separation of a pair of entangled particles produced by the quantum vacuum occurs, one particle falling into the black hole and the other moving away from the event horizon into space. It has been proposed~\cite{Polchinski} that the information attached to the infalling particle is not lost but is ``memorized'' by the outgoing particle. The sharing of information in the radiation process generates a firewall near the horizon. However, the local recognition of the existence of the firewall event horizon goes against the cherished notion that an infalling observer in GR does not know he/she is falling through an event horizon (equivalence principle) and this creates a paradox.   

The stochastic quantization of the spacetime metric leads to a probabilistic interpretation of spacetime, based on defining the metric tensor as a random stochastic variable~\cite{Moffat}. A similar approach to quantizing the gravitational field was proposed by others~\cite{Damgaard,Rumpf}. It has been demonstrated that statistical stochastic quantization is equivalent to quantum field quantization in Euclidean space~\cite{Parisi}, and this equivalence has been extended to a spacetime with a Lorentz signature~\cite{HuffelRumpf}. A canonical formalism based on a $(3+1)$ foliation of spacetime in GR leads to a Langevin equation~\cite{Langevin} for the metric tensor treated as a stochastic variable. 
Classical GR is based on the assumption that the spacetime manifold is $C^2$ smooth down to zero length scales. The singularity theorems of Hawking and Penrose~\cite{Penrose}, which state that in GR the final state of gravitational collapse results in a singularity and a black hole event horizon, are based on this assumption. The assumption that spacetime is $C^2$ smooth at all length scales contradicts our knowlege of physical systems in Nature, such as those occurring in physics, chemistry and biology. We know that physical systems display noise as random fluctuations, $1/f^n$ flicker noise or chaotic noise at some length scale.  When we consider spacetime as a physical system that interacts with matter, then we expect that the spacetime metric will also experience fluctuations correponding to quantum ''noise`` in the spacetime geometry. The assumption that spacetime geometry is smooth to zero length scales appears to be quite unrealistic.

\section{Spacetime Stochastic Equations and Probability Densities}

The diffeomorphism invariant Einstein field
equations including stochastic spacetime fluctuations take the form~\cite{Moffat}:
\begin{equation}
G_{\mu\nu}=8\pi G(T_{\mu\nu}+\sigma\xi_xT_{\mu\nu}),
\end{equation}
where $G_{\mu\nu}$ is the Einstein tensor: $G_{\mu\nu}=R_{\mu\nu}-\frac{1}{2}g_{\mu\nu}R$.
Moreover, $\xi_x$ denotes Gaussian quantum spacetime fluctuations and $\sigma$ is a measure of the strength of the fluctations.

We introduce a $(3+1)$ foliation of spacetime which defines a 3-geometry ${}^{(3)}{\cal G}$, and choose a time
function $t$ and a vector field $t^\mu$ such that the surfaces, $\Sigma$, of constant
$t$ are spacelike Cauchy surfaces with $t^\mu\nabla_\mu t=1$, and where $\nabla_\mu$
denotes the covariant differentiation with respect to the metric tensor $g_{\mu\nu}$~\cite{Arnowitt,Wald}. 

The canonically conjugate momentum variable $\pi^{\mu\nu}$ is defined by
\begin{equation}
\pi^{\mu\nu}=\frac{\partial {\cal L}_G}{\partial (\partial_th_{\mu\nu})}
=\sqrt{h}(K^{\mu\nu}-h^{\mu\nu}K),
\end{equation}
where $K_{\mu\nu}$ is the extrinsic curvature and $K=h^{\mu\nu}K_{\mu\nu}$. Moreover, we have 
\begin{equation}
{\cal L}={\cal L}_G+{\cal L}_M, 
$$ $$
{\cal L}_G=\sqrt{-g}g^{\mu\nu}R_{\mu\nu},
\end{equation}
and ${\cal L}_M$ is the Lagrangian density for the matter field. We have $g={\rm det}\,g_{\mu\nu}$, 
$R_{\mu\nu}$ is the Ricci curvature tensor, $h_{\mu\nu}=g_{\mu\nu}+n_\mu n_\nu$ is the induced spatial metric and $n^\mu$ is the normal unit vector to the Cauchy spacelike surface, $\Sigma$. 

The stochastic momentum variable $\pi^{\mu\nu}_t$ obeys a stochastic differerential equation (SDE)~\cite{Moffat}:
\begin{equation}
\partial_t\pi^{\mu\nu}_t=F_t^{\mu\nu}+8\pi G(T^{\mu\nu}+\sigma\xi_tT^{\mu\nu}).
\end{equation}
Here, $F_t^{\mu\nu}$ involves the stochastic $(3+1)$ projected curvature tensor, $\pi^{\mu\nu}_t$ and the lapse and shift functions, $T^{\mu\nu}$ is the projected stress-energy tensor for matter with the components, $T_{\perp\perp}=T_{\mu\nu}n^\mu n^\nu$, $T_\perp^\nu={h^\nu}_\alpha T^{\alpha\beta}n_\beta$.
For a given 3-geometry ${}^{(3)}{\cal G}$, including geometrical quantum noise in the
gravitational equations leads to the formal Langevin~\cite{Langevin} or SDE:
\begin{equation}
\label{SDE}
\partial_tX_t(g_t)=f(g_t)+\sigma\xi_th(g_t).
\end{equation}
The extreme irregularity of Gaussian white noise means that the time derivative of the
metric is not well defined. However, the standard theory of stochastic processes can
handle this difficulty by defining the equivalent integral equation:
\begin{equation}
X_t(g_t)=X_0+\sigma\int^t_0h(g_y)\xi_ydy.
\end{equation}
Two definitions of the stochastic integral $\int h(g_y)\xi_ydy$ have been given by
Ito~\cite{Ito} and Stratonovich~\cite{Stratonovich}.

The probability density $p(y,t\vert g,\tau)$ of the metric diffusion
process $g_t$ satisfies the Fokker-Planck equation (FPE)~\cite{Fokker,Planck} or Kolmogorov forward equation:
\begin{equation}
\label{Fokkereq}
\partial_tp(y,t\vert g,\tau)=-\partial_y[f(y,t)p(y,t\vert g,\tau)]
+\frac{1}{2}\partial_{yy}[h^2(y,t)p(y,t\vert g,\tau)].
\end{equation}
The functional derivatives occurring in (\ref{Fokkereq}) exist and
are continuous and for convenience we have suppressed tensor indices.
The FPE equation for time homogeneous Markov metric processes, is given by
\begin{equation}
\partial_tp(y,t\vert g)=-\partial_y[f(y)p(y,t\vert g)]+\frac{1}{2}\partial_{yy}
[h^2(y)p(y,t\vert g)],
\end{equation}
where the drift and the diffusion are time independent. For a one-dimensional system, the
FPE for the probability density becomes
\begin{equation}
\partial_tp(g,t)=-[\partial_gf(g,t)p(g,t)]+\frac{1}{2}\partial_{gg}[h^2(g,t)p(g,t)].
\end{equation}

When the boundaries of the gravitational system are natural, i.e., regular with
instantaneous reflection imposed as a boundary condition, then there is no
flow of probability out of the state space and by using the Ito
prescription~\cite{Ito}, we obtain~\cite{Horsthemke,Gihman,Oksendal}:
\begin{equation}
\label{stochasticprobability}
p_S(g)=\frac{C}{h^2(g)}\exp\biggl(\frac{2}{\sigma^2}\int^g
\frac{f(x)}{h^2(x)}dx\biggr),
\end{equation}
where $C$ is a normalization constant. In this case, the normalization constant $C$ is given by
\begin{equation}
C^{-1}=\int_{g_1}^{g_2}\frac{1}{h^2(g)}\exp\biggl(\frac{2}{\sigma^2}
\int^g\frac{f(x)}{h^2(x)}dx\biggr)dg <\infty.
\end{equation}

\section{Test Particle Infall into a Black Hole}

For the geodesic motion of a test particle, a stochastic equation is given by
\begin{equation}
du^\mu_s+\Gamma^\mu_{s,\alpha\beta}u_s^\alpha
u_s^\beta+\sigma\xi_s\Gamma^\mu_{s,\alpha\beta}u^\alpha_s u^\beta_s ds=0,
\end{equation}
where $\xi_s$ is a spacetime quantum fluctuation process in terms of the proper time $s$, the Christoffel
symbol $\Gamma^\mu_{s,\alpha\beta}$ is treated as a random variable determined by the
stochastic metric $g_{s,\mu\nu}$, $u^\mu=dx^\mu/ds$ denotes the time-like four-velocity,
and $u_s^\mu$ describes this four-velocity as a random variable.
For big enough macroscopic length scales for which the spacetime fluctuations can be neglected, the motion
of the test particle becomes the same as the deterministic geodesic equation of motion in GR.

A stochastic Raychaudhuri equation can be derived~\cite{Moffat} from which we can deduce
that for a big enough intensity of fluctuations caustic singularities in the spacetime manifold can be prevented from occurring if 
the standard positive energy conditions in GR are valid. In the limit of classical GR, the Hawking-Penrose singularity theorems will continue
to hold, for the fluctuations of spacetime are negligible and can be neglected.

Consider the case of inward radial motion in a Schwarzschild geometry. We have
\begin{equation}
{dt\over ds}=\biggl(1-{2GM\over r}\biggr)^{-1},\quad\quad {dr\over ds}
=-\sqrt{2GM\over r},
\end{equation}
where $M$ denotes the mass of the central particle. The Langevin equation for the
random variable $r_s$ is given by
\begin{equation}
dr_s=f(r_s)ds+\sigma\xi_sf(r_s)ds,
\end{equation}
where 
\begin{equation}
f(r_s)=-\sqrt{2GM\over r_s}
\end{equation}
By taking the
white-noise Gaussian limit for short correlation times~\cite{Horsthemke,Gihman,Oksendal,Moffat}, we get using (\ref{stochasticprobability}) the stationary probability density:
\begin{equation}
p_S(r)={Cr\over 2M}\exp\biggl(-{2\sqrt{2}\sqrt{G}\over 3\sqrt{M}\sigma^2}r^{3/2}\biggr).
\end{equation}
Both the drift and the diffusion coefficients vanish as $r\rightarrow\infty$, for $r\rightarrow\infty$ is a natural boundary.  
We have $p_S(r)\sim 0$ in the limit $r\rightarrow 0$ and we conclude that as the particle
falls towards the origin there is zero probability for $r(s)$ to have the value zero. Consequently
there is zero probability density of having a singularity at $r=0$ in the Schwarzschild solution, because the
spacetime metric fluctuations smear out the singularity at $r=0$. 

\section{Stochastic Collapse of a Dust Cloud}

We apply stochastic quantum gravity to the gravitational collapse of a star~\cite{Moffat}. We will follow the Oppenheimer-Snyder~\cite{Oppenheimer,Weinberg} treatment of the collapse. 
The metric in co-moving coordinates describing a freely falling dust cloud is given by
\begin{equation}
ds^2=dt^2-U(r,t)dr^2-V(r,t)(d\theta^2+\sin^2\theta d\phi^2).
\end{equation}
The energy-momentum tensor is
\begin{equation}
T^{\mu\nu}=\rho u^\mu u^\nu,
\end{equation}
where $\rho(r,t)$ is the proper energy density and $u^\mu$ is 
given in comoving coordinates by
\begin{equation}
u^r=u^{\theta}=u^{\phi}=0,\quad u^0=1.
\end{equation}

We can seek a separable solution for homogeneous collapse:
\begin{equation}
U(r,t)=R^2(t)f(r),\quad V(r,t)=S^2(t)g(r).
\end{equation}
Einstein's field equations require that ${\dot S}/S={\dot R}/R$,
where ${\dot R}=\partial_tR$ and we can choose : $S(t)=R(t)$ and redefine the 
radial coordinate, so that $V=R^2(t)r^2$.  The GR field equations lead to the Friedmann-Robertson-Walker (FRW) metric:
\begin{equation}
\label{FRW}
ds^2=dt^2-R^2(t)\biggl[\frac{dr^2}{1-kr^2}+r^2d\theta^2
+r^2\sin^2\theta d\phi^2\biggr].
\end{equation}

We choose $R(t)$ so that $R(0)=1$ and we get
\begin{equation}
\rho(t)=\rho(0)R^{-3}(t).
\end{equation}
The field equations yield
\begin{equation}
\label{Rdoteq}
{\dot R}^2(t)=\frac{8\pi G}{3}\frac{\rho(0)}{R(t)}-k,
\end{equation}
where $k$ is a constant given by
\begin{equation}
k=\frac{8\pi G}{3}\rho(0).
\end{equation}

By choosing an integrating factor, we can transform the metric to a standard
form~\cite{Weinberg}:
\begin{equation}
ds^2=B(r,t)dt^2-A(r,t)dr^2-r^2(d\theta^2+\sin^2\theta d\phi^2),
\end{equation}
where
\begin{eqnarray}
B&=&\frac{R}{W}\biggl(\frac{1-kr^2}{1-ka^2}\biggr)^{1/2}
\frac{(1-ka^2/W)^2}{(1-kr^2/R)},\\
A&=&\biggl(1-\frac{kr^2}{R}\biggr)^{-1}.
\end{eqnarray}
The constant $a$ is equated to the radius of the collapsing sphere in comoving polar coordinates and we have 
\begin{equation}
W(r,t)=1-\biggl(\frac{1-kr^2}{1-ka^2}\biggr).
\end{equation}
It follows that the interior and the exterior solutions match continuously at $r=aR(t)$
when $k=2GM/a^3$, which gives $M=\frac{4\pi}{3}\rho(0)a^3$. The first order equation of motion is
\begin{equation}
\label{sqrtdotR}
{\dot R}(t)=-\biggl(\frac{2GM}{a^3}\biggr)^{1/2}\biggl[\frac{1}{R(t)}-1\biggr]^{1/2},
\end{equation}
where for the collapse problem we have chosen the negative square root. The stochastic Langevin equation for $R_t$ is given by
\begin{equation}
dR_t=f(R_t)dt+\sigma\xi_tf(R_t)dt,
\end{equation}
where
\begin{equation}
\label{ffunction}
f(R_t)=-\biggl(\frac{2GM}{a^3}\biggr)^{1/2}\biggl(\frac{1}{R_t}-1\biggr)^{1/2}.
\end{equation}

In the white-noise limit of rapidly varying spacetime fluctuations, we obtain the approximate stationary probability density:
\begin{equation}
p_S(R)=\frac{C}{f^2(R)}\exp\biggl[\frac{2}{\sigma^2}\int^R\frac{dR}{f(R)}\biggr].
\end{equation}
From (\ref{ffunction}) we find
\begin{equation}
\label{probdensity}
p_S(R)\sim C\biggl(\frac{a^3}{2GM}\biggr)
\biggl(\frac{R}{1-R}\biggr)
\exp\biggl\{\frac{2}{\sigma^2}\biggl(\frac{a^3}{2GM}\biggr)^{1/2}
[\sqrt{R}\sqrt{1-R}-\arcsin{\sqrt{R}}]\biggr\}.
\end{equation}
We arrive at the result that $p_S(R)\sim 0$ as $R(t)\rightarrow 0$ and there is a zero probability for $R(t)$ to have the value zero. {\it There is a zero probability 
of having a singularity as the final state of collapse.}

The fractional change of wave length emitted at the surface is
\begin{equation}
z=\frac{\lambda^\prime-\lambda_0}{\lambda_0}
=\frac{dt^\prime}{dt}-a{\dot R}(t)\biggl(1-\frac{2GM}{aR(t)}\biggr)^{-1}-1,
\end{equation}
where
\begin{equation}
t^\prime=t+\int^t_{aR(t)}\biggl(1-\frac{2GM}{r}\biggr)^{-1}dr
\end{equation}
is the time it takes for a light signal emitted in a radial direction at standard time $t$
to reach a distant point $r$.  In the limit, $R(t)\rightarrow 2GM/a=ka^2$, we obtain
by using (\ref{sqrtdotR}):
\begin{equation}
\label{zeq}
z\sim2\biggl(1-\frac{ka^2}{R(t)}\biggr)^{-1}.
\end{equation}
We have
\begin{equation}
dz=f(R)dt,
\end{equation}
where
\begin{eqnarray}
f(R)&=& -\frac{4GM}{a}\biggl(1-\frac{2GM}{aR}\biggr)^{-2}
\frac{{\dot R}}{R^2}\nonumber\\
&=&\biggl(\frac{2}{a}\biggr)^{5/2}(GM)^{3/2}\biggl(1-\frac{2GM}
{aR}\biggr)^{-2}
\frac{(1-R)^{1/2}}{R^{5/2}}.
\end{eqnarray}

We can now define the Langevin SDE:
\begin{equation}
dz_t=f(R_t)dt+\sigma\xi_tf(R_t)dt,
\end{equation}
where the function $f(R)\rightarrow\infty$ as $R\rightarrow 2GM/a$, corresponding
to the infinite red shift limit as the event horizon is approached during collapse
predicted by classical GR, and $\xi_t$ is a short time stochastic quantum process. In the white-noise limit the stationary probability density is given by
\begin{equation}
\label{probability}
p_S(z)=\frac{C\biggl(1-\frac{2GM}{aR}\biggr)^4R^5}{N^2(1-R)}\exp\bigg[\frac{2}{\sigma^2N}\int\frac{dR\biggl(1-\frac{2GM}{aR}\biggr)^2R^{5/2}}{(1-R)^{1/2}}\biggr],
\end{equation}
where $N=(2/a)^{5/2}(GM)^{3/2}$.
Thus, in the limit $R\rightarrow 2GM/a$ the probability for a black hole event horizon with an infinite redshift to form during gravitational collapse is zero. On the other hand, once a 
trapped surface forms in {\it classical GR collapse}, the redshift seen by a distant observer must inevitably become infinite at the radius $r_S=2GM$. 

The geometrical fluctuations in stochastic quantum gravity about the classical deterministic event horizon cut
off the high wave length at a finite value, $\lambda=\lambda_c$, as viewed by an
observer at large distances from the collapsing star. In fact, the distant observer does not see a
collapsed object described by a classical static Schwarzschild metric with no hair. The collapsed object
will fluctuate about the average classical Schwarzschild solution with a characteristic time
determined by the size of the correlation function for the Schwarzschild metric.
The red shift of the collapsed object can be high, so that an outside observer would
believe that it is a black hole. The cooperative effects associated with the
self-organising microscopic subsystems comprising the event horizon can produce a large or
even infinite correlation length for the microscopic fluctuations, cutting off the infinite
wavelength radiation emitted by the macroscopic surface of the collapsing star as $R\rightarrow 2GM/a$.

The spacetime fluctuations as $R\rightarrow 2GM/a$ quench the infinite red shifts. 
The standard GR results follow when we can ignore the spacetime metric fluctuations: both the classical singularity and the infinite red shift
black hole horizon must occur when a trapped surface forms during collapse. However, this
assumes that the spacetime geometry is perfectly smooth in the limit of zero distance scales and
infinite frequencies of red shifts, an assumption {\it which seems contrary to all our experiences of
physical systems.} The Hawking-Penrose results for classical GR collapse are not valid for large enough fluctuations, because caustic singularities in the metric are prevented from occurring for converging congruences of timelike or null geodesics in the spacetime manifold~\cite{Moffat}. The stationarity of the probability density assumes that after an infinite time has elapsed the collapsing star has evolved to a stationary equilibrium state. 
This is compatible with the spherically symmetric collapse possessing a well-defined asymptotically flat space limit as $r\rightarrow\infty.$ In gravitational collapse, it can take an infinite time for a distant observer to see a black hole event horizon form, which is consistent with the observer measuring the stationary probability density $p_S$ as $t\rightarrow\infty$. 

Because the probability of a black hole event horizon forming in stochastic quantum gravity is zero, the concept of the entropy of a stable black hole event horizon with the Bekenstein-Hawking entropy:
$S_{\rm BH}=A/4G\hbar$, where $A$ is the area of the horizon, has to be reconsidered. The quantum spacetime fluctuations of the area $A=4\pi R^2$ will affect the standard ideas of holographic and thermodynamic physics of black holes.

\section{Conclusions}

The zero probability of black hole event horizons forming means that there are no black holes in the classical sense of GR from which no light can escape to infinity. 
However, trapped surfaces can form during collapse that can produce apparent horizons, which appear as a grey surface to a distant observer. The greyness of the collapsed object depends on the size
of the fluctuating correlation time $\tau_{\rm cor}$. The fluctuations occur around the classical mean value of the apparent horizon, so that there is a leakage from the grey hole and this solves the
Hawking information loss paradox~\cite{Hawking,Hawking2,Hawking3}. The spectrum of the radiation emitted from the surface of the grey apparent horizon will be modified from the purely random lack of information behavior of Hawking radiation associated with a classical GR event horizon. Moreover, the lack of a black hole event horizon would prevent the formation of a firewall.

\section{Acknowledgements}

The John Templeton Foundation is thanked for its generous support of
this research. The research was also supported by the Perimeter
Institute for Theoretical Physics. The Perimeter Institute was
supported by the Government of Canada through Industry Canada and by
the Province of Ontario through the Ministry of Economic Development
and Innovation. We also would like to thank Martin Green and Viktor Toth for helpful discussions and comments.

\end{document}